# BERRY PHASE AND TRAVERSAL TIME IN ASYMMETRIC GRAPHENE STRUCTURES

D. Dragoman

Univ. Bucharest, Physics Dept., P.O. Box MG-11, 077125 Bucharest, Romania

**Abstract:**

The Berry phase and the group-velocity-based traversal time have been calculated for an asymmetric non-contacted or contacted graphene structure, and significant differences have been observed compared to semiconductor heterostructures. These differences are related to the specific, Dirac-like evolution law of charge carriers in graphene, which introduces a new type of asymmetry. When contacted with electrodes, the symmetry of the Dirac equation is broken by the Schrödinger-type electrons in contacts, so that the Berry phase and traversal time behavior in contacted and non-contacted graphene differ significantly.

**Introduction**

The Berry phase [1] is encountered in all areas of quantum or classical optical physics. It is the phase acquired by a system subjected to cyclic adiabatic processes and which originates in the geometrical properties of the parameter space of the Hamiltonian [2, 3], or the phase acquired during propagation through a multilayer system [4]. In graphene, the quantum Berry phase in the presence of a magnetic field leads to the anomalous half-integer quantum Hall effect [5,6], which is caused by the specific topology of the band structure near the Dirac point [7]. In asymmetric semiconductor heterostructures described either by the Schrödinger equation [8] or by the two-band Kane model [9], the Berry phase associated with transport through a multilayer system is linked to different tunneling times for the opposite directions of propagation, which correspond to the same transmission probability. In these works, a group-velocity-based tunneling time was used. A similar definition of traversal time in graphene has been recently introduced in [10], based on the similarity between the Dirac equation and the two-band Kane model [11]. Unlike in structures with Schrödinger-type ballistic charge carriers, in which the only asymmetry is that with respect to the direction of propagation, in a gated graphene sheet there is an additional asymmetry, with respect to the sign of the incidence angle of charge carriers. The last type of asymmetry is related to the fact that the Dirac Hamiltonian transforms into its complex conjugate when the incidence angle changes sign, whereas the Schrödinger equation is not affected by this sign change. In [10], only normally incident charge carriers on a gated graphene region with or without contacts have been considered, case in which the eventual time asymmetry related to the incidence angle could not have been observed, and no attempt was made to evidence the existence of time asymmetry associated to opposite propagation directions.

In this paper, we study the effect of these two asymmetry types on the argument of the amplitude transmission coefficient, which can be considered as a Berry phase, and on the traversal time. We consider graphene sheets without and with electrodes. In the first case, the traversal time and Berry phase are not affected by the sign of the incidence angle, but sense only the asymmetry with respect to the propagation direction, whereas in the second case both asymmetry types are observed in the traversal time. The reason is that the presence of electrodes, in which charge carriers propagate according to the Schrödinger equation, breaks the symmetry of the Dirac equation with respect to the change in sign of the incidence angle.

**Berry phase and traversal time in gated graphene without electrodes**

In graphene, the low-energy charge carriers satisfy the Dirac equation. Let us consider an asymmetric structure consisting of three regions with different potential energies $V_j$, $j = 1, 2, 3$, the central one extending from $x = 0$ to $x = D$. The potential energies can be modified by doping or gating. In each region $j$, the Dirac equation is

$$\hbar v_F \begin{pmatrix} 0 & k_j - ik_y \\ k_j + ik_y & 0 \end{pmatrix} \begin{pmatrix} \psi_1 \\ \psi_2 \end{pmatrix} = (E - V_j) \begin{pmatrix} \psi_1 \\ \psi_2 \end{pmatrix} \tag{1}$$

where $v_F \cong c/300$ is the Fermi velocity (with $c$ the speed of light), $k_y = k_{Fj} \sin \varphi_j$ and $k_j = k_{Fj} \cos \varphi_j = [(E - V_j)^2 / \hbar^2 v_F^2 - k_y^2]^{1/2}$ are the components of the Fermi wavenumber $k_{Fj} = (E - V_j)/\hbar v_F$ along the $y$ and $x$ directions, respectively, and $E$ is the energy of charge carriers. According to the boundary conditions, $k_y$ is unchanged at propagation. The solution of (1) is commonly written as

$$\psi_1(x, y) = \exp(ik_y y) \times \begin{cases} \exp(ik_1 x) + r \exp(-ik_1 x), & x \leq 0 \\ a \exp(ik_2 x) + b \exp(-ik_2 x), & 0 < x < D \\ t \exp(ik_3 x), & x \geq D \end{cases} \tag{2a}$$

$$\psi_2(x,y) = \exp(ik_y y) \times \begin{cases} s_1[\exp(ik_1 x + i\varphi_1) - r\exp(-ik_1 x - i\varphi_1)], & x \leq 0 \\ s_2[a\exp(ik_2 x + i\varphi_2) - b\exp(-ik_2 x - i\varphi_2)], & 0 < x < D \\ s_3 t \exp(ik_3 x + i\varphi_3), & x \geq D \end{cases} \quad (2b)$$

where $\varphi_j = \tan^{-1}(k_y/k_j)$, $j = 1, 2, 3$, $s_j = \text{sgn}(E - V_j)$, $r$ and $t$ are the amplitude reflection and transmission coefficients, respectively, and $a$ and $b$ are determined from the wavefunction continuity requirement at the $x = 0$ and $x = D$ interfaces. The transmission probability through the structure is defined as $T = s_3 \cos(\varphi_3)|t|^2 / s_1 \cos(\varphi_1)$ and, similar to type II/III heterostructures [10], we define a group-velocity-based traversal time for charge carriers propagating along the $x$ direction as

$$\tau = \int_0^D \frac{dx}{v_g(x)} = \int_0^D \frac{\rho(x)dx}{J}. \quad (3)$$

Here $v_g(x) = J/\rho(x)$ is the group velocity in graphene expressed in terms of the probability density $\rho = |\psi_1|^2 + |\psi_2|^2$ and the probability current along $x$, $J = v_F(\psi_1 \psi_2^* + \psi_1^* \psi_2)$.

To asses the influence of both types of asymmetry on the traversal time and on the Berry-like phase $\vartheta$, defined through $t = |t|\exp(i\vartheta)$, we focus first on normal incidence on the asymmetric structure. After a straightforward calculus, from (2) we obtain $t = 2s_1 s_2 \exp(-ik_3 D)/[s_2(s_1 + s_3)\cos(k_2 D) - i(1 + s_1 s_3)\sin(k_2 D)]$, which implies that $t$ is not defined if $s_1 \neq s_3$, i.e. if the charge carriers in the incident and output media (regions 1 and 3) are different: electrons versus holes. Moreover, irrespective of the asymmetry (the different $V_j$ values), $\tau/\tau_0$ with $\tau_0 = D/v_F$ equals unity when the emerging charge carriers are electrons and is equal to $-1$ when holes are collected at the output. A non-singular value of $t$ can be always obtained if the Dirac spinors are written as

$$\psi_1(x,y) = \exp(ik_y y) \times \begin{cases} \exp(ik'_1 x) + r\exp(-ik'_1 x), & x \leq 0 \\ a\exp(ik'_2 x) + b\exp(-ik'_2 x), & 0 < x < D \\ t\exp(ik'_3 x), & x \geq D \end{cases} \quad (4a)$$

$$\psi_2(x,y) = \exp(ik_y y) \times \begin{cases} \exp(ik'_1 x + i\varphi'_1) - r\exp(-ik'_1 x - i\varphi'_1), & x \leq 0 \\ a\exp(ik'_2 x + i\varphi'_2) - b\exp(-ik'_2 x - i\varphi'_2), & 0 < x < D \\ t\exp(ik'_3 x + i\varphi'_3), & x \geq D \end{cases} \quad (4b)$$

with $k'_j = s_j k_j$, $\varphi'_j = s_j \varphi_j$. The latter form was shown to avoid the Klein paradox at a step barrier [12]. In the latter case, $T = \cos(\varphi_3)|t|^2 / \cos(\varphi_1)$, and at normal incidence $\vartheta = (k'_2 - k'_3)D$ when the structure is traversed from region 1 to 3, and $\vartheta = (k'_2 - k'_1)D$ if traversed from region 3 to 1, $\tau/\tau_0$ being equal to unity in all cases, irrespective of the potential energy values. The expressions obtained from the two formula, (2) and (4), are identical if the incident and output charge carriers are electrons.

These analytical results show that the direction-of-propagation asymmetry does not influence the traversal time in graphene at normal incidence, conclusion which is opposite to the results obtained for semiconductor heterostructures [8,9], for which a clear asymmetry is detected in the traversal/tunnelling time at normal incidence, and this asymmetry is strongly dependent on $D$ and the energy potential values. Moreover, this time asymmetry is related to the phase difference of $r$ for the two propagation directions.

Further, we consider a symmetric structure ($V_1 = V_3$) and look for the effect of incidence angle sign on $T$, $\vartheta$ and $\tau/\tau_0$. The results are presented in Figs. 1(a) and 1(b) for $V_1 = V_3 = 0.05$ eV, $D = 50$ nm, $E = 0.15$ eV and $V_2 = 0.2$ eV (solid line) and 0.25 eV (dashed line). Throughout this paper quantities corresponding to the same set of parameters are represented with the same line type. As expected, $T$ does not depend on the sign of $\varphi_1$, and neither do the other parameters; at normal incidence $T = 1$, irrespective of the values of other parameters. The sign of the incidence angle does not matter even for asymmetric structures, with $V_1 \neq V_3$, as can be seen from Figs. 2(a) and 2(b), in which the same plots are made for $V_1 = 0$, $V_2 = 0.2$ eV, and identical values for the other parameters. The left-to-right and right-to-left Berry phases (black) and traversal times (gray) are displayed with solid and dashed lines, respectively. In asymmetric structures different directions of propagation lead to different $\vartheta$ and $\tau/\tau_0$ but the same transmission probability.

To conclude this section, the direction-of-propagation asymmetry is always associated to changes in the Berry phase, is linked with changes in the traversal time only for obliquely incident charge carriers, and does not influence the transmission probability $T$. We will see that this behaviour is not maintained in contacted graphene. From the simulations in Figs. 1 and 2 it follows that the normalized traversal time is significantly different from unity in the immediate neighbourhood of regions where charge carrier transport is forbidden due to the lack of an energy band gap in graphene (see also [10] for a detailed discussion of this subject), and follows the change in Berry phase in the sense that the traversal time is larger when the Berry phase is steeper.

**Berry phase and traversal time in gated graphene with electrodes**

When the charge carriers originate from and are collected by electrodes, where they satisfy a Schrödinger equation, a mathematical method must be devised to describe the transformation of Schrödinger electrons into Dirac electrons in graphene and vice-versa, which takes place at the electrode/graphene interface. In this section we consider a gated graphene region, labeled by 2, which extends from $x = 0$ to $x = D$, bordered by semi-infinite source and drain regions labeled by 1 and 3, respectively. The electron wavefunction in the source and drain at oblique incidence is a scalar, given by

$$\psi = \exp(ik_y y) \times \begin{cases} \exp(ik_1 x) + r\exp(-ik_1 x), & x \leq 0 \\ t\exp(ik_3 x), & x \geq D \end{cases} \qquad (2)$$

where $k_j = [2m_j(E-V_j)/\hbar^2]^{1/2} \cos\varphi_j$, $j = 1, 3$, and $m_j$ is the effective electron mass in the metallic electrodes, while in the gated graphene channel, for $0 < x < D$, we have, as in the previous section, two possible expressions. However, since both incoming and outgoing charge carriers are electrons, there is no difference between the results obtained using these two expressions. Therefore, we write the wavefunction in region 2 as

$$\psi = \begin{pmatrix} \psi_1 \\ \psi_2 \end{pmatrix} = \exp(ik_y y) \times \begin{pmatrix} a\exp(ik'_2 x) + b\exp(-ik'_2 x) \\ a\exp(ik'_2 x + i\varphi'_2) - b\exp(-ik'_2 x - i\varphi'_2) \end{pmatrix} \qquad (3)$$

with $k_y = [2m_1(E-V_1)/\hbar^2]^{1/2} \sin\varphi_1 = [2m_3(E-V_3)/\hbar^2]^{1/2} \sin\varphi_3 = k_{F2} \sin\varphi_2$. The only way to impose boundary conditions between a scalar and a spinor wavefunction is to take advantage of the formal similarity between $\psi_1$ and the wavefunction in the 1 and 3 regions, and between $\psi_2$ and the $x$ derivative normalized to $m_j$ of the wavefunctions in regions 1 and 3. This similarity is justified if we regard $a$ and $b$ as amplitudes of forward and backward plane wave components. Then, we impose the boundary conditions of continuity of similar functions on each side of the interface,

$$1 + r = a + b, \qquad \frac{k_1}{m_1}(1-r) = \frac{v_F}{\hbar}(ae^{i\varphi'_2} - be^{-i\varphi'_2}), \qquad (4a)$$

$$ae^{ik'_2 L} + be^{-ik'_2 L} = te^{ik_3 L}, \qquad \frac{v_F}{\hbar}(ae^{ik'_2 L + i\varphi'_2} - be^{-ik'_2 L - i\varphi'_2}) = \frac{k_3}{m_3} te^{ik_3 L}, \qquad (4b)$$

the spinor component $\psi_2$ being multiplied with the constant $v_F/\hbar$ due to dimensionality considerations. It can easily be checked that the boundary conditions (4) also warrant the conservation of the current probability across the structure, in electrodes this parameter being defined as $J = (\hbar/2mi)[\psi^*(d\psi/dx) - \psi(d\psi^*/dx)]$. The traversal time in graphene is defined as in the previous section, while the transmission probability is now given be $T = (k_3 m_1 / k_1 m_3)|t|^2$.

In graphene contacted with ohmic metallic electrodes, we have in general $V_1 \cong V_3 \cong 0$ and $m_1 \cong m_3 \cong m_0$, with $m_0$ the free electron mass. However, for the sake of argument we consider first an asymmetric structure, for which $V_1 = 0$, $V_2 = 0.25$ eV, $V_3 = 0.05$ eV, $m_1 = m_3 = m_0$ and $D = 20$ nm. As expected, at normal incidence the transmission probability $T$ does not depend on the propagation direction and the Berry phase for left to right propagation (solid line) differs from the value of the same parameter if the structure is traversed from right to left (dashed line). The dependence of these parameters on $E$ is shown in Fig. 3. However, unlike in the previous section, we find that the traversal times (gray, with the same line type as the corresponding Berry phase) for the two propagation direction are no longer the same. The reason is that the Schrödinger equation valid for charge carriers in electrodes and the Dirac equation obeyed by charge carriers in graphene have a different behavior as the sign of the incidence angle changes. Therefore, the electrodes break the symmetry of charge carriers in graphene with respect to the incidence angle sign. Note that at normal incidence the transmission probability $T$ is no longer equal to 1, due to the mismatch between electrodes and graphene.

The traversal time asymmetry becomes more pronounced for obliquely incident electrons, when the two asymmetry types combine. For example, the normalized difference between the left-to-right and right-to-left traversal times, denoted as $\Delta\tau$, increases as the asymmetry (the value of $V_3$) in the above example increases and the angle of incidence increases, as seen from Fig. 4 where the solid, dotted and dashed lines correspond to incidence angles of 0, 2 and 4 degrees; the traversal time asymmetry for normal incidence is evident also in this figure. Note that for symmetrical structures with respect to the direction of propagation (with $V_1 = V_3$ and $m_1 = m_3$), the left to right and right to left traversal times are identical, irrespective of the angle of incidence, i.e. the angle-of-incidence related asymmetry is manifest only in structures asymmetric with respect to the direction of propagation. In Fig. 5 we have shown the dependence on the incident angle of the transmission and of the normalized $\Delta\tau$ for $V_1 = 0$, $V_3 = 0.05$ eV, $E = 0.15$ eV, $D = 20$ nm, and for three values of $V_2$: 0.2 eV (dashed line), 0.25 eV (solid line), and 0.3 eV (dotted line). The corresponding curves for $T$ and $\Delta\tau$ are shown with the same line type in Figs. 5(a) and 5(b), respectively. The angle of incidence dependence of the difference of Berry phases for the two propagation direction is much weaker and was not plotted (the difference of Berry phases for the opposite propagation directions is almost constant and equal to 1 rad for the example chosen). Note the strong angle asymmetry of $\Delta\tau$, and the independence of $T$ on the sign of $\varphi_1$. The same angle asymmetry of $\Delta\tau$ is obtained also if the energy $E$ is varied (not shown). From Fig. 5 it follows that $\Delta\tau$ can take larger or smaller values for obliquely incident electrons than those for normal incidence. Moreover, unlike in uncontacted graphene sheets, where both the Berry phase and the traversal time depend on the magnitude but not sign of the incidence angle, in contacted graphene the dependence of $\vartheta$ on $\varphi_1$ is much weaker than the dependence of the traversal time. Both parameters depend also on the sign of the incidence angle and can thus be tuned by changing this sign. The simulations presented in this paper also confirm the close connection between the Berry phase and the band-structure topology.

**Conclusions**

We have studied the dependence of the Berry phase and of the group-velocity-based traversal time in asymmetric, contacted and non-contacted graphene structures and found significant differences in behavior compared to semiconductor structures described by the Schrödinger or two-band Kane models. First, in graphene an additional asymmetry type can be identified, which is related to the sign of the incidence angle. Second, in non-contacted graphene the Berry phase and traversal time do not depend on the incidence angle sign, and the traversal time is different for the two propagation directions only for oblique charge carries. This behavior changes dramatically in contacted graphene, where both types of asymmetries can be observed, because the presence of electrodes breaks the symmetry of the Dirac equation in graphene with respect to the change of sign of the incidence angle. The fact that the traversal time can be changed by modifying the incidence angle can be used to tune the value of this parameter, which becomes particularly large in the neighborhood of regions where charge carrier propagation is forbidden by the absence of a band gap in graphene.

**Figure Captions**

Fig. 1  Incidence angle dependence of (a) transmission probability, and (b) Berry phase (black) and normalized traversal time (gray) for $V_1 = V_3 = 0.05$ eV, $D = 50$ nm, $E = 0.15$ eV and $V_2 = 0.2$ eV (solid line) and 0.25 eV (dashed line).

Fig. 2  Incidence angle dependence of (a) transmission probability, and (b) Berry phase (black) and normalized traversal time (gray) for the same structure as in Fig. 1 but with $V_1 = 0$, and $V_2 = 0.2$ eV. The left-to-right and right-to-left parameters are displayed with solid and dashed lines, respectively.

Fig. 3  Energy dependence of the (a) transmission probability, and (b) Berry phase (black) and normalized traversal time (gray) for contacted graphene with $V_1 = 0$, $V_2 = 0.25$ eV, $V_3 = 0.05$ eV, $m_1 = m_3 = m_0$ and $D = 20$ nm. The left-to-right and right-to-left parameters are displayed with solid and dashed lines, respectively.

Fig. 4  Asymmetry dependence of the difference in traversal times for the opposite propagation directions for the system in Fig. 3 with $E = 0.15$ eV. The solid, dotted and dashed lines correspond to incidence angles of 0, 2 and 4 degrees.

Fig. 5  Incidence angle dependence of the (a) transmission and (b) normalized $\Delta\tau$ for $V_1 = 0$, $V_3 = 0.05$ eV, $E = 0.15$ eV, $D = 20$ nm, and $V_2 = 0.2$ eV (dashed line), 0.25 eV (solid line), and 0.3 eV (dotted line).

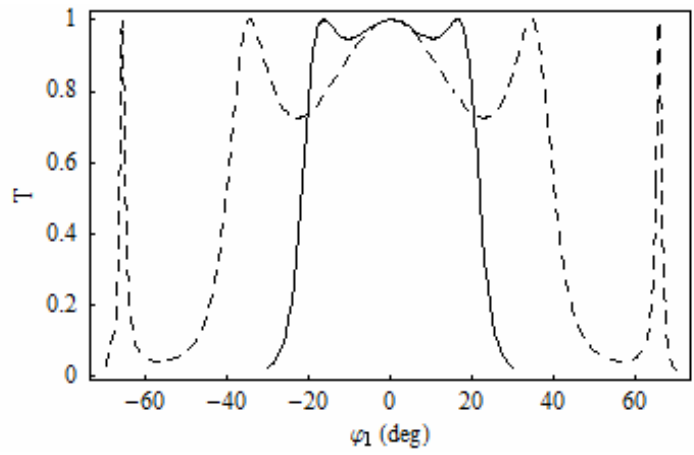

(a)

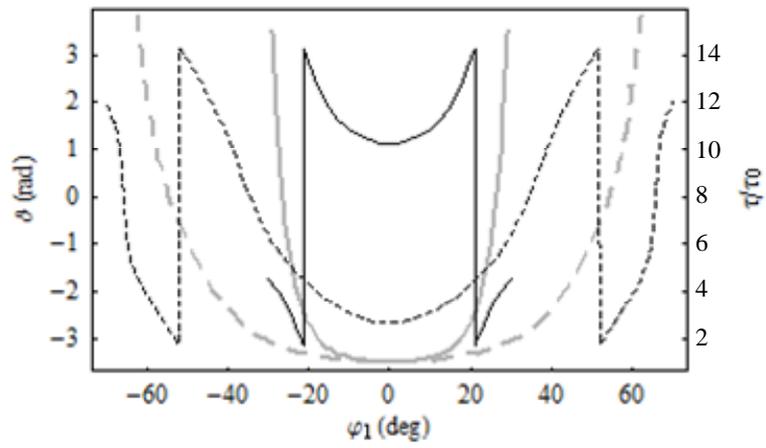

(b)

Fig. 1

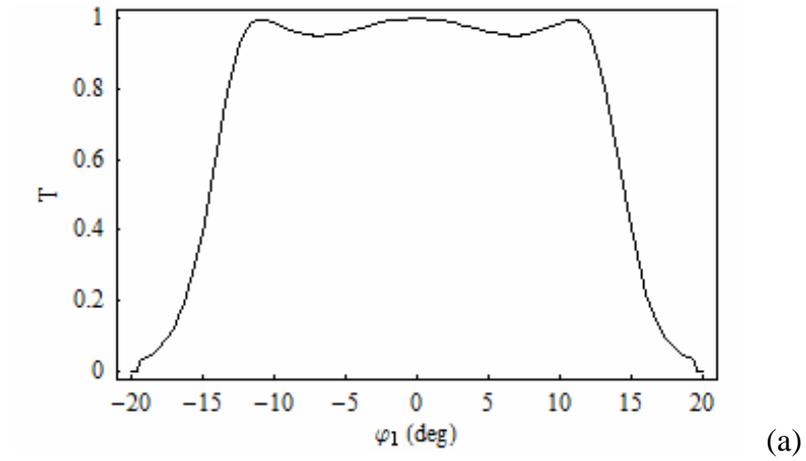

(a)

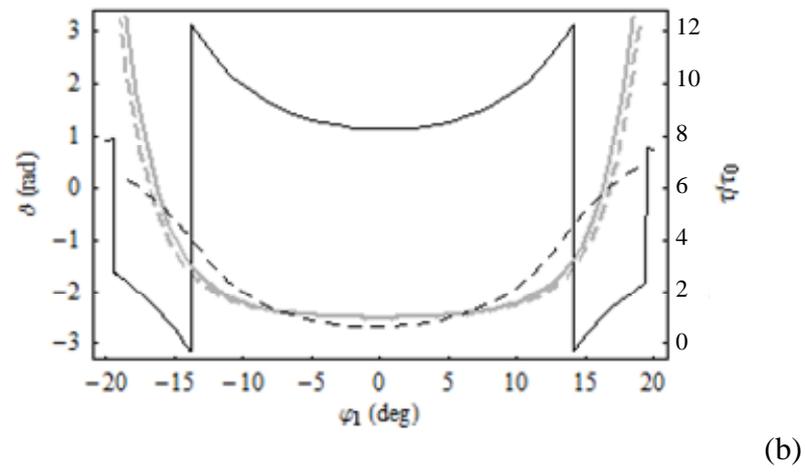

(b)

Fig. 2

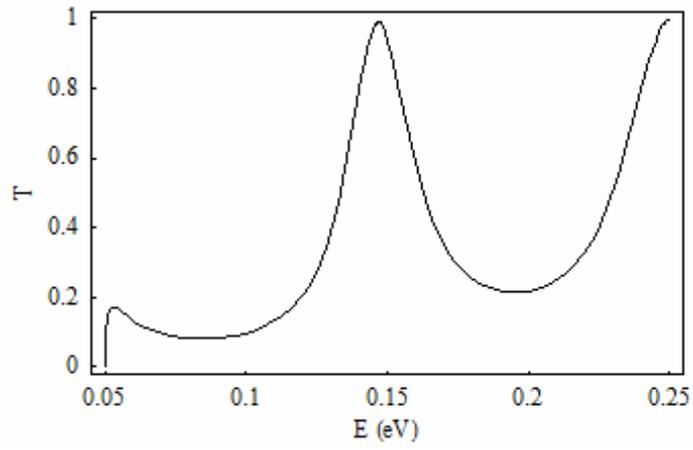

(a)

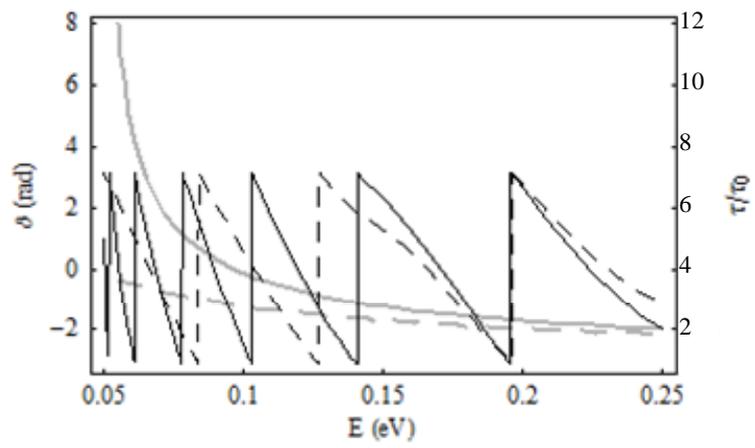

(b)

Fig. 3

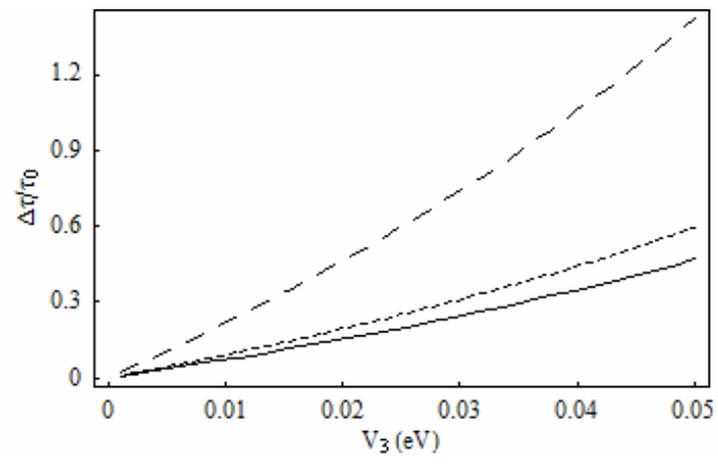

Fig. 4

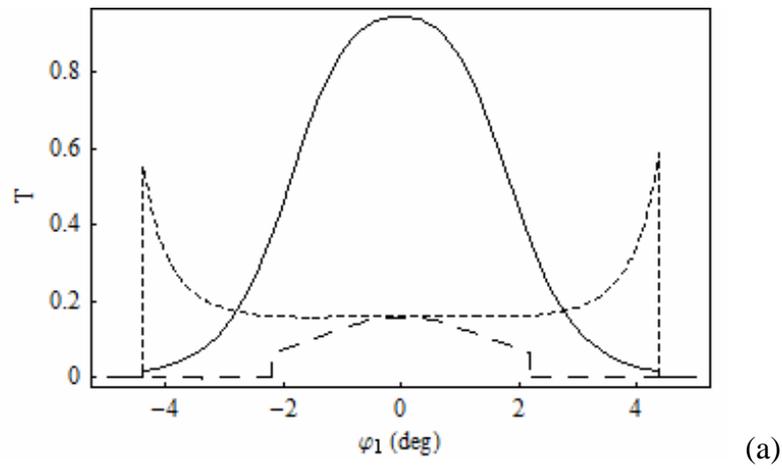

(a)

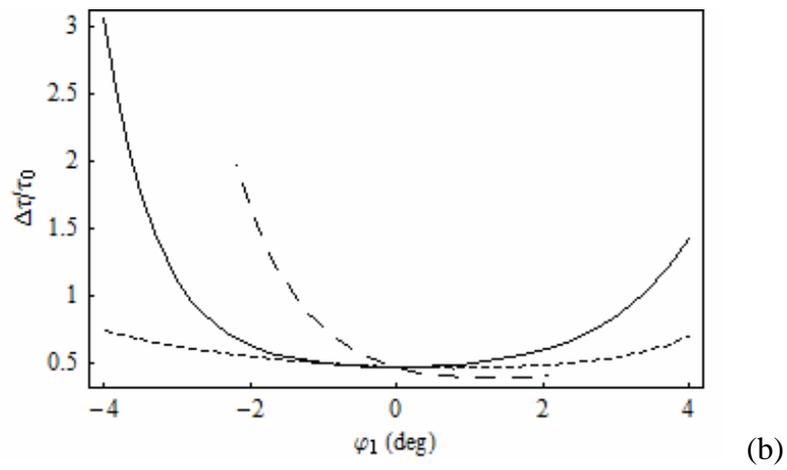

(b)

Fig. 5